\documentclass[twocolumn,twoside,slac_two]{revtex4}
\usepackage{graphicx}
\usepackage{fancyhdr}
\usepackage{hyperref}
\hypersetup{
    colorlinks=true,
    urlcolor=blue,
}
\usepackage{eurosym}

\usepackage[percent]{overpic}

\catcode`\@=11 
\def\parenbar{\mathpalette\p@renb@r}
\def\p@renb@r#1#2{\vbox{%
  \ifx#1\scriptscriptstyle \dimen@.7em\dimen@ii.2em\else
  \ifx#1\scriptstyle \dimen@.8em\dimen@ii.25em\else
  \dimen@1em\dimen@ii.4em\fi\fi \offinterlineskip
  \ialign{\hfill##\hfill\cr
    \vbox{\hrule width\dimen@ii}\cr
    \noalign{\vskip-.3ex}%
    \hbox to\dimen@{$\mathchar300\hfil\mathchar301$}\cr
    \noalign{\vskip-.3ex}%
    $#1#2$\cr}}}
\catcode`\@=12 
\def\nuan{\parenbar{\nu}\kern-0.4ex}

\begin{document}

\title{Prospects for measuring the neutrino mass hierarchy with KM3NeT/ORCA}

\author{J.~Hofest{\"a}dt on behalf of the KM3NeT Collaboration}
\affiliation{
Erlangen Centre for Astroparticle Physics,
Friedrich-Alexander University of Erlangen-N\"urnberg,\\ 
Erwin-Rommel-Str.~1, 91058 Erlangen, Germany
}

\begin{abstract}
ORCA (Oscillation Research with Cosmics in the Abyss) is the low-energy branch of KM3NeT, 
the next-generation research infrastructure hosting underwater Cherenkov detectors in the Mediterranean Sea.
ORCA's primary goal is the determination of the neutrino mass hierarchy
by measuring the matter-induced modifications on the oscillation probabilities of few-GeV atmospheric neutrinos.
The ORCA detector design foresees a dense configuration of KM3NeT neutrino detection technology, 
optimised for measuring the interactions of neutrinos in the energy range of $3$--$20$\,GeV.
To be deployed at the French KM3NeT site, 
ORCA's multi-PMT optical modules will exploit the excellent optical properties of deep-sea water to accurately reconstruct 
both shower-like (mostly electron neutrino) and track-like (mostly muon neutrino) events
in order to collect a high-statistics sample of few-GeV neutrino events.

This contribution reviews the methods and technology of the ORCA detector, 
and discusses the prospects for measuring the neutrino mass hierarchy
as well as the potential to improve the measurement precision on other oscillation parameters.
\end{abstract}

\maketitle

\thispagestyle{fancy}

\section{Introduction}
\label{sec:intro}
A variety of experiments with solar, atmospheric, reactor and accelerator neutrinos, 
spanning energies from MeV up to tens of GeV, 
demonstrated unambiguously that neutrinos change from one flavour to another 
during propagation.
Neutrino oscillations imply non-zero neutrino masses,
and that the masses of the three neutrino states are different.
In the standard $3$-neutrino scheme, 
the mixing matrix relating the neutrino flavour eigenstates ($\nu_e$, $\nu_\mu$, $\nu_\mu$)
to the mass eigenstates ($\nu_1$, $\nu_2$, $\nu_3$) 
is parameterised in terms of three mixing angles $\theta_{12}$, $\theta_{13}$ and $\theta_{23}$,
and a CP-violating phase $\delta_{\text{CP}}$.
Oscillation experiments are mostly sensitive to mass-squared differences $\Delta m_{ij}^2 = m^2_i - m^2_j ~ (i,j=1,2,3)$.
Global fits of available data form a coherent picture and provide the values of these oscillation parameters with reasonable precision \cite{globalFit}.

An open question is the so-called neutrino mass hierarchy (NMH). 
It refers to the ordering of the neutrino mass eigenstates,
which is either 
$\nu_1<\nu_2<\nu_3$ (normal hierarchy, NH) 
or $\nu_3<\nu_1<\nu_2$ (inverted hierarchy, IH). 
The ordering of the first two closely spaced mass eigenstates,
$\nu_1 < \nu_2$,
is known from solar neutrino physics.
Further yet unknown neutrino properties are:
the value of $\delta_{\text{CP}}$,
the absolute masses, 
the Dirac/Majorana nature of neutrinos.
Knowing the NMH is important for constraining the models that seek to explain the origin of mass in the leptonic sector
and will allow to optimise the information obtained from other neutrino experiments
(targeting $\delta_{\text{CP}}$, absolute neutrino masses and neutrinoless double-beta decays).
In addition, the NMH has a significant impact on the measurement precision of the oscillation parameters.

The NMH can be determined by measuring the energy and zenith angle dependent
oscillation pattern of few-GeV atmospheric neutrinos that have traversed the Earth towards the detector \cite{AkhmedovRazzaqueSmirnov}. 
Due to matter-induced modifications on the oscillation probabilities
in conjunction with different cross-sections and atmospheric neutrino fluxes for neutrinos and antineutrinos, 
the expected event rates of neutrinos in the energy regime of $3$--$20$\,GeV are different for NH and IH.

Next-generation experiments, such as KM3NeT/ORCA \cite{KM3NeT_LoI}, PINGU \cite{PINGU} and Hyper-Kamiokande \cite{HyperK},
are planned to perform this measurement with megaton-scale water/ice-based Cherenkov detectors.

In the following, the prospects for measuring the neutrino mass hierarchy with 
ORCA (Oscillation Research with Cosmics in the Abyss)
are presented,
and the potential to improve the measurement precision on $\theta_{23}$ and $\Delta m_{32}^2$ is discussed.

This contribution is mainly based on the `Letter of Intent for KM3NeT 2.0' \cite{KM3NeT_LoI}.

\section{The KM3NeT/ORCA detector}
\label{sec:detector}
The KM3NeT detector design builds on the experience of the successful deployment and operation of the ANTARES detector \cite{ANTARES},
which has demonstrated the feasibility of measuring neutrinos with a large-volume Cherenkov detector in the deep sea.
The detection principle is that of a 3-dimensional array of photosensors 
that register the Cherenkov light induced by charged particles
produced in a neutrino-induced interaction. 
From the arrival time of the Cherenkov photons (nanosecond precision) 
and the position of the sensors ($\sim$10\,cm precision), 
the energy and direction of the incoming neutrino, 
as well as other parameters of the neutrino interaction, 
can be reconstructed.

\begin{figure}
\includegraphics[width=55mm]{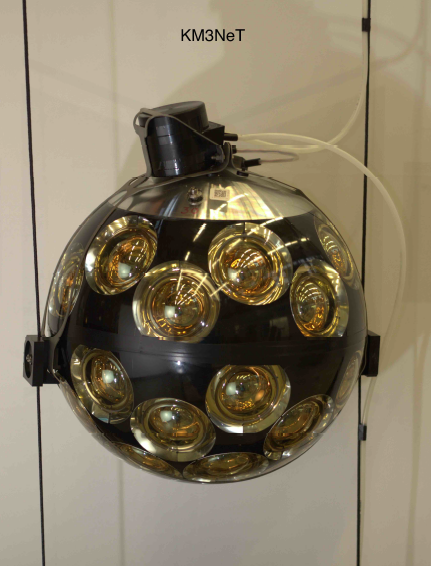}
\hfill
\includegraphics[width=21.7mm]{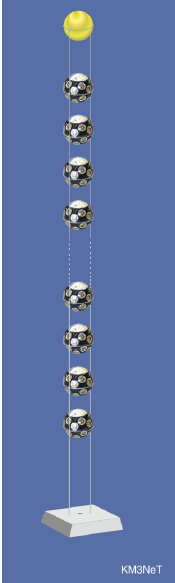}
\caption{Photograph of a DOM (left) and schematic drawing of an detection string (right).}
\label{fig:DOM_string}
\end{figure}

A key KM3NeT technology is the Digital Optical Module (DOM),
a pressure-resistant glass sphere housing 31 3-inch PMTs and their associated electronics.
This multi-PMT design offers several improvements compared to 
traditional optical modules hosting only a single large PMT (for example in ANTARES),
most notably: larger photocathode area, wider field of view, directional information and dynamic range.
The DOMs are arranged in strings
held vertically by a buoy and anchored to the seabed.
Figure~\ref{fig:DOM_string} shows a DOM and a detection string.

In its current design,
the ORCA detector will comprise $115$ such detection strings.
Each string comprises $18$ DOMs with a vertical spacing of about $9$\,m.
The horizontal spacing between adjacent strings is roughly $20$\,m.
The instrumented mass is about $6$\,Mton of seawater.
This detector configuration is the outcome of a optimisation study
using the NMH sensitivity as figure of merit.
The proposed detector could be built in three years, 
with an investment budget of about $45\,$M\euro\ \cite{procPaschalNeutrino16}.

The ORCA detector will be deployed at the French KM3NeT site at a depth of $2450$\,m.
The site is about $40$\,km offshore from Toulon and about $10$\,km west of the operating ANTARES detector.

The construction of the infrastructure 
has already started.
The first main electro-optical cable
and the first junction box, needed to connect the detection strings,
have been successfully deployed and connected.
The first detection string is foreseen to be deployed in early 2017.
An array comprising 7 detection strings is funded and expected to be concluded and operational by the end of 2017. 
It will serve to demonstrate the feasibility of the measurement and to validate and optimise the detector design.
The full-size ORCA detector comprising $115$ detection strings could be operational towards 2020.

Within KM3NeT,
the same technology is employed also for 
the search for high-energy astrophysical neutrino sources with the ARCA detector \cite{ARCA_these_proceedings},
which will be deployed offshore from Sicily, Italy.
The main difference between both detector designs is the density
of photosensors, which is optimised for the study of neutrinos in the few-GeV (ORCA) and
TeV-PeV (ARCA) energy ranges.

\section{Expected Detector Performance}
\label{sec:detector_performance}

The key parameters for the NMH determination are the effective mass of the detector 
and the experimental resolutions for the energy $E_\nu$ and zenith angle $\theta_\nu$ of the incoming neutrino. 

Detailed Monte Carlo simulations have been performed 
using GENIE\,\cite{genie} for simulating neutrino interactions 
and GEANT-based simulation packages 
\cite{hours,antSimTools}
for particle propagation and Cherenkov photon generation. 
Optical background from ${}^{40}$K decays in the seawater 
as well as the background from downgoing atmospheric muons
is taken into account.
Further details are given in \cite{KM3NeT_LoI}.

Two distinct event topologies are considered: tracks and showers. 
Showers are initiated by energetic electrons and hadrons emerging from the neutrino interaction,
and develop over relatively short distances.
Muons produce elongated tracks in the detector.
Therefore, track-like events are induced by $\nuan_\mu$ charged-current (CC) interactions, 
as well as $\nuan_\tau$~CC interactions with muonic tau decays.
All other neutrino-induced events are called shower-like, 
i.e. $\nuan_e$~CC events, $\nuan_{e,\mu,\tau}$ neutral-current events 
and $\nuan_\tau$~CC events with non-muonic $\tau$ decays.

Dedicated reconstruction strategies for track-like and shower-like events,
as well as an event topology classification algorithm,
have been developed and are described in \cite{KM3NeT_LoI}.
The energy resolution is Gaussian-like with $\sigma_{E_\nu} / E_\nu \approx 25$\%.
The median zenith angle resolution is about $5^\circ$
for $\nuan_e$~CC and $\nuan_\mu$~CC events with $E_\nu=10$\,GeV.
Due to the long scattering length of light in deep-sea water,
the reconstructions are able to find the lepton ($e,\mu$) in $\nuan_{e,\mu}$~CC events 
and are therefore able to gain access to the interaction inelasticity $y$.
This allows a statistical separation of $\nu$ and $\overline{\nu}$ interactions to further add to the NMH sensitivity \cite{RibordySmirnov}. 
This possibility has not yet been exploited in the estimated NMH sensitivity presented below.
As shown in \cite{intrinsicLimits},
the energy resolution is dominated by intrinsic light yield fluctuations in the hadronic shower
and the direction resolution is limited by the kinematic scattering angle between the outgoing lepton and the incoming neutrino.

The purity of the event topology classification is about 90\% (70\%) 
for $\nuan_e$~CC ($\nuan_\mu$~CC) events with $E_\nu=10$\,GeV.
The same event classification algorithm also rejects downgoing atmospheric muons 
that are mis-reconstructed as upgoing.
A contamination of less than a few percent of atmospheric muons
in the final sample of upgoing neutrino events
is achieved.

\begin{figure}
{\begin{overpic}[width=85mm]{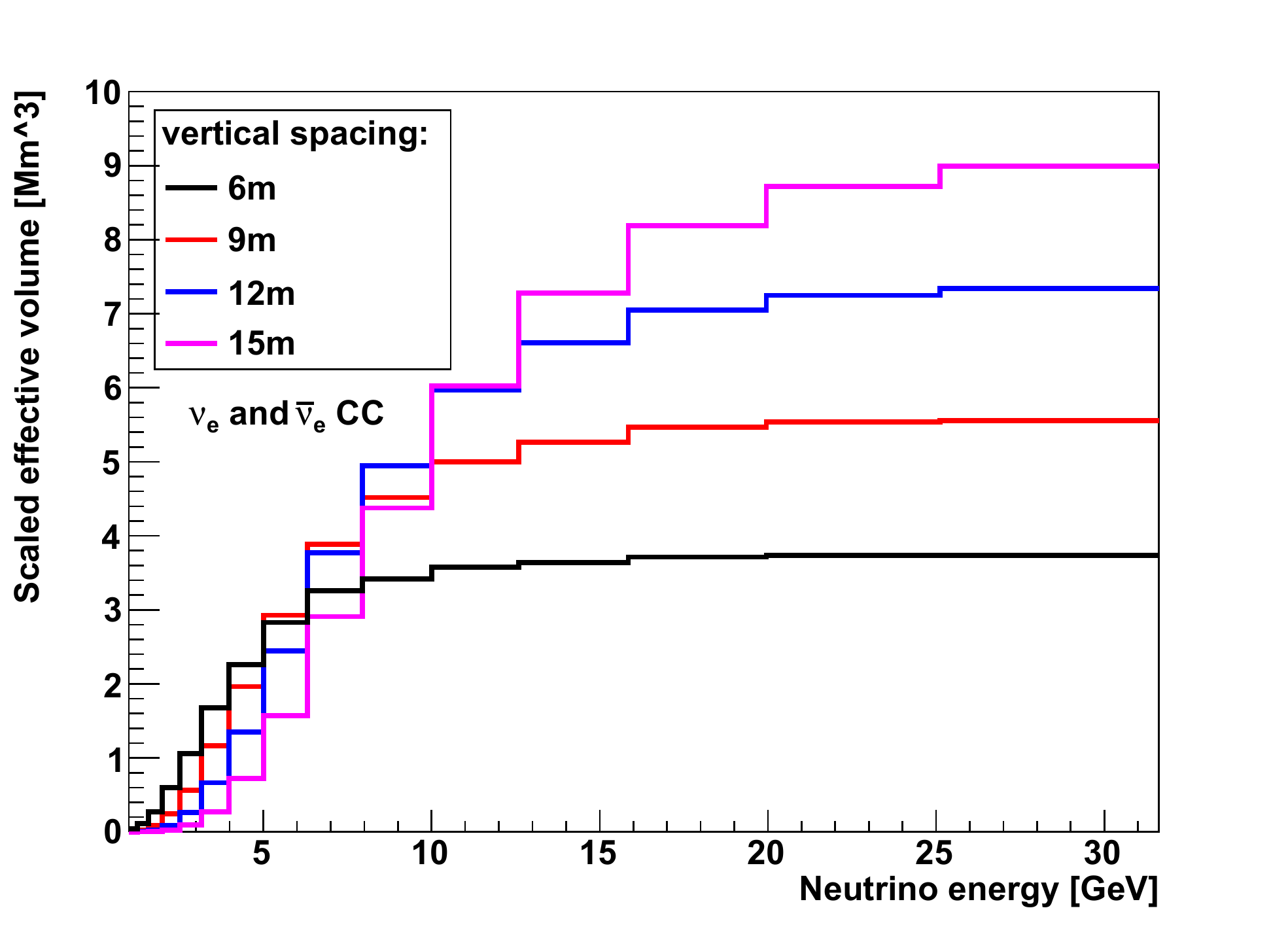} 
\put (37,60) {\bf KM3NeT}
\end{overpic}}
\caption{Effective volume as a function of neutrino energy $E_\nu$ for $\protect \nuan_e$~CC events. 
Detector configurations with different vertical spacings between the DOMs are shown as different colours.}
\label{fig:effMass_nue_diff_vspacings}
\end{figure}

The effective mass of the detector is about $6$\,Mton,
being reached for $\nuan_e$~CC and $\nuan_\mu$~CC with energies above $E_\nu=10$\,GeV,
while 50\% efficient at $4$\,GeV.
Figure~\ref{fig:effMass_nue_diff_vspacings} shows the effective volume 
for $\nuan_e$~CC events for detector configurations with different vertical spacings between the DOMs,
i.e.\ different photosensor densities.
In the $E_\nu=5-10$\,GeV range,
which is most relevant for the NMH determination,
the detector configuration with $9$\,m vertical spacing provides the largest effective mass 
and therefore largest available event statistics.
This detector configuration was also found to provide the best NMH sensitivity \cite{KM3NeT_LoI}.
It will provide data samples of about 50,000 reconstructed upgoing neutrinos per year. 

\section{Sensitivity to neutrino mass hierarchy and more}
\label{sec:sensitivity}

Building on the expected detector performance,
a significance analysis is performed by generating a large number of 
pseudo-experiments (PEs)
with event distributions in the reconstructed $E_\nu$--$\theta_\nu$ plane. 
For each PE, a true hierarchy and a set of oscillation parameters is assumed.
Each PE is analysed by performing a maximum likelihood fit with the oscillation parameters as free parameters and assuming either NH or IH.
The likelihood ratio resulting from these fits is used to quantify the 
separability between both hierarchies.

Systematic uncertainties from the neutrinos fluxes, their cross sections as well as the detector response
are parameterised as overall normalisation, energy scale, $\overline{\nu}$/$\nu$ skew, $\mu$/$e$ skew and NC/CC skew
and are incorporated as nuisance parameters.
It is found that none of these effects compromise substantially the ability of ORCA to determine the NMH \cite{KM3NeT_LoI}. 

\begin{figure}
\includegraphics[width=80mm]{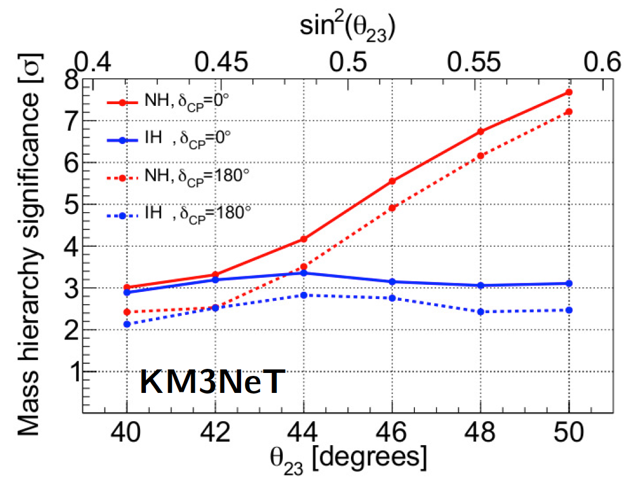}
\caption{Median NMH significance to exclude the other hierarchy hypothesis assuming true NH (red) or IH (blue)
as a function of true $\theta_{23}$ and assuming $\delta_{\protect \text{CP}} = 0$ (solid) or $\delta_{\protect \text{CP}} = \pi$ (dashed).
Three years of data taking with the full-size ORCA detector are assumed.}
\label{fig:NMH_significance}
\end{figure}

Figure~\ref{fig:NMH_significance} shows the median significance of ORCA 
to exclude the wrong hierarchy hypothesis
after three years of data taking as a function of the true value of $\theta_{23}$ 
and assuming no CP-violation, i.e. $\delta_{\text{CP}}$ equals $0$ or $\pi$.
For the experimentally allowed range of $\theta_{23}$ and assuming $\delta_{\text{CP}} = 0$, 
the NMH can be measured with about $3\,\sigma$ after three years of operation. 
ORCA is moderately sensitive to the CP-phase, 
the significance being reduced by at most $0.5\,\sigma$
if $\delta_{\text{CP}} = \pi$ is realised in nature.
The significance increases dramatically
in case of NH and $\theta_{23} > \pi/4$,
reaching up to about $7\,\sigma$ in three years of operation.

\begin{figure}
\includegraphics[width=75mm]{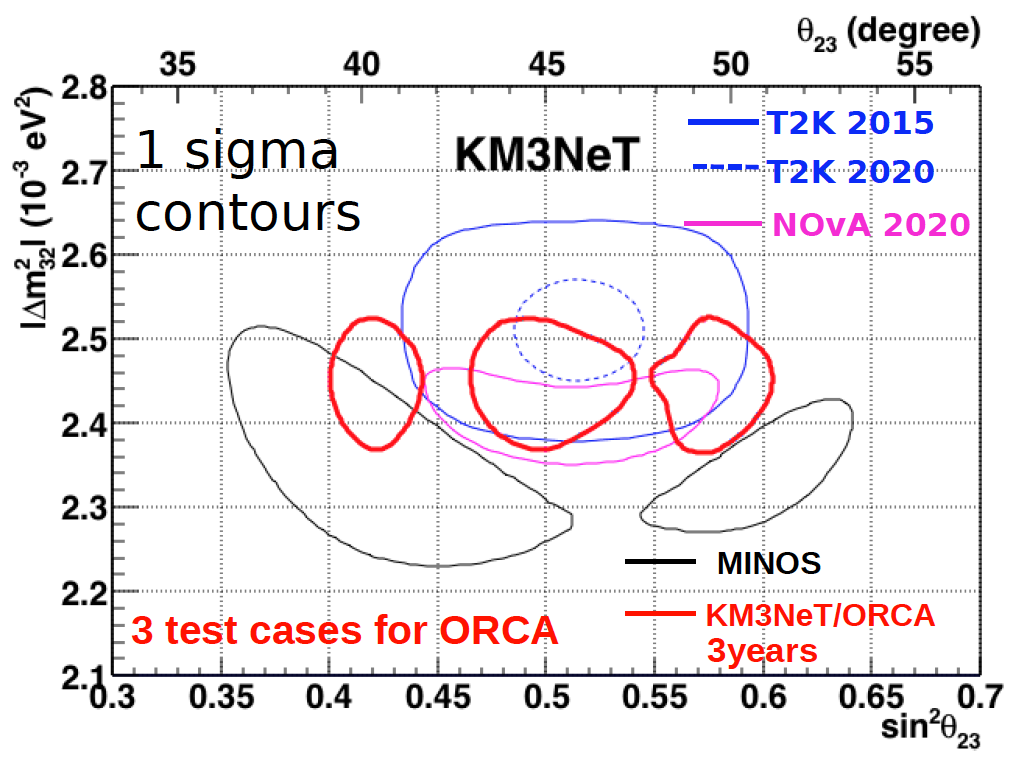}
\caption{One sigma contours of the measurement precision in $\Delta m_{32}^2$ and $\theta_{23}$
after three years of data taking with ORCA for three 
assumed 
test cases (red).
The current results from \mbox{MINOS} (back) and T2K (blue solid) are indicated, 
as well as the predicted performance of NO$\nu$A (magenta) and T2K (blue dashed) in 2020. 
All contours are at $1\,\sigma$ for NH.}
\label{fig:osci_para}
\end{figure}

Besides the NMH determination, 
ORCA can also improve the uncertainties on
$\Delta m_{32}^2$ and $\theta_{23}$.
Both parameters are determined 
without the need for constrains from global data
in conjunction with the NMH. 
Figure~\ref{fig:osci_para} shows the expected 
measurement precision
after three years
and compares it with current results of other experiments and their predicted performances in 2020.
The precision of ORCA is comparable or better, 
and is obtained 
with different systematic uncertainties.
In particular,
ORCA can determine the ‘octant’ of $\theta_{23}$
(above or below $45^\circ$)
for a wide range of the
allowed parameter range.

Additional science topics of ORCA include:
testing the unitary of the neutrino mixing matrix by studying $\nuan_\tau$ appearance; 
indirect searches for sterile neutrinos, non-standard interactions and other exotic physics; 
indirect searches for dark matter; 
testing the chemical composition of the Earth's core (Earth tomography);
and low-energy neutrino astrophysics.
Preliminary performance expectations are briefly summarised in \cite{procPaschalNeutrino16}.
The KM3NeT research infrastructure will also house instrumentation for
Earth and Sea sciences, such as marine biology, oceanography and geophysics.

Possible future options could be a long-baseline neutrino beam targeted to ORCA \cite{Brunner},
and a significantly denser detector instrumentation lowering the detection threshold 
to measure the CP-phase $\delta_{\text{CP}}$ with atmospheric neutrinos \cite{RazzaqueSmirnov_CP}.

\section{Conclusions}
\label{sec:conclusions}

With ORCA, 
a $6$\,Mton deep-sea Cherenkov detector optimised for the detection of few-GeV neutrinos,
the KM3NeT Collaboration aims to perform a high-statistics measurement 
of the zenith angle and energy dependent event rates of atmospheric neutrinos 
that have traversed the Earth.
The oscillated neutrino flux in the energy range $3$--$20$\,GeV
holds the key to determine the neutrino mass hierarchy.
ORCA is expected to achieve a $3$--$7\,\sigma$ sensitivity to the neutrino mass hierarchy in three years of data taking.
Simultaneously, ORCA will measure $\Delta m_{32}^2$ and $\theta_{23}$
with competitive precision,
and has a rich additional science program.

In the first construction phase of ORCA, 
a 7-string demonstrator is expected to be concluded and operational by the end of 2017.
The full-size ORCA detector could be operational towards 2020,
so that the neutrino mass hierarchy might be resolved as early as 2023.

Further details can be found in the recently published 
`Letter of Intent for KM3NeT 2.0' \cite{KM3NeT_LoI}.

\bigskip 

\end{document}